\documentclass[a4paper]{article}
\usepackage{graphicx}

\begin{document}

\begin{center}
{\bf \Large
A note on temperature without energy - a social example}
\bigskip

{\large
K. Ku{\l}akowski$^{\dag}$
}
\bigskip

{\em
Faculty of Physics and Applied Computer Science,
AGH University of Science and Technology,
al. Mickiewicza 30, PL-30059 Krak\'ow, Poland\\

}

\bigskip

$^\dag${\tt kulakowski@fis.agh.edu.pl}

\bigskip
%\today
\end{center}

The question if the statistical mechanics can be useful in social sciences is open.
Physicists who try to apply it there - the sociophysicists \cite{book,santo}
- are often enthusiastic, including this author. As reported in \cite{mywind}, reactions of the social scientists are much more
critical. Here we are interested to what
extent the concept of temperature can be directly applied in sociology. In the statistical
mechanics, temperature is a measure of the mean kinetic energy of particles. However,
there is no direct equivalent of energy in social sciences. We expect intuitively that at
high temperatures there is more noise and disorder. Actually, this rough interpretation
motivated the use of the term 'social temperature' in classical texts \cite{stauf}.
Often, temperature is introduced as a disorder parameter in model formulas borrowed directly
from statistical physics \cite{wio,kaho}.\\

In a recent preprint \cite{lmf}, temperature of a social system was introduced
as the derivative of the entropy density with respect to the amount of cooperation.
This way is promising, as the entropy can be defined for any probability distribution
and the concept of energy is not needed. However, the problem appears what condition
should be imposed to the quantity with respect to which the derivative is calculated.
Using energy, volume or number of particles we get $1/T$, $p/T$ or $\mu /T$, respectively,
where $T$ is the temperature, $p$ is the pressure and $\mu$ is the chemical potential. Then
it is not clear how $T$ is entangled in the derivative used in Ref. \cite{lmf}. On the
other hand, the paper \cite{lmf} is full of original, new and interesting concepts (as
the Dipole model), which are however not necessary if we want merely to show that temperature
can be defined quantitatively in a social system. Such a demonstration - as simple as possible -
is our aim here. The note presented here is motivated by Ref. \cite{lmf} and profits from
this reference.\\

We consider a Bethe lattice of $N$ agents, each with three neighbours, where each agent plays one of two strategies A or B; according to the 
standard notation, orientations $s_i=\pm 1$ are assigned to these strategies. Denoting by $n$
the number of those who play A, their proportion to the whole set is $p=n/N$. This $p$ is
assumed to vary in time due to three processes: {\it i)} averaging over nearest neighbours, {\it ii)} random flipping, {\it iii)} following mass media, which suggest A.
The respective rates are $w$, $q$ and $h$. Let us denote $2p-1$ by $m$. The equation of motion 
is 

\begin{equation}
\frac{dm}{dt}=\frac{wm}{2}(1-m^2)-qm+\frac{h}{2}(1-m)
\end{equation}
The first term on the r.h.s of this equation is a sum of the contributions of two configurations: when an agent playing A, surrounded by three agents playing B, converts to B, and the oposite: an agent playing B, surrounded by three agents playing A, converts to A. These contributions are $-p(1-p)^3$ and $p^3(1-p)$, respectively. The contributions from other configurations are zero or mutually cancel. The term with $q$ is proportional to $-m$ because a random flip reduces the average
orientation. With time $t$ rescaled, the equation becomes

\begin{equation}
\frac{dm}{d\tau}=\frac{m}{2}(1-m^2)-\frac{y}{2}m+ \frac{x}{2}(1-m)
\end{equation}
where $\tau=wt$, $y=2q/w$ and $x=h/w$. At zero field, we have three fixed points: $m=0$ (always exists, stable iff $y>1$) and $m=\pm \sqrt{1-y}$, which exist and are stable iff $y<1$. For the purposes of this text we can limit ourselves to the case $y>1$. Switching on the influence of media
with $x$ infinitesimally small, we can define the 'zero-field' susceptibility

\begin{equation}
\chi=\frac{dm}{dx}
\end{equation}
For $y>1$ we get 

\begin{equation}
\chi=\frac{1}{y-1}
\end{equation}
which is analogous to the Curie-Weiss law \cite{magnb}

\begin{equation}
\chi=\frac{1}{T-T_C}
\end{equation}
where $T$ is the temperature. and $T_C$ is its critical value, where the spontaneous magnetization is different from zero. We recognize $y$ - the rate of random changes of strategies - as a social equivalent of temperature. Actually, 'strategy' can be reinterpreted as 'opinion' or any other kind of binary characteristics of an agent.\\

Concluding, the concept of susceptibility offers a simple way of the quantitative interpretation
of the social temperature. A question remains, if this concept - temperature as randomness - will 
be qualified as useful by the social scientists. This seems possible, if the sociophysicists are able 
to show how in human life the randomness spreads from one event to another through a chain of consequences,
in a way similar to the equalization of temperature in physics.

\end{document}